\documentstyle[12pt]{article}
 at 14.4truept at 12.0truept
\title{ The Gravity dual of the Non-Perturbative $ N = 2 $ SUSY
Yang-Mills Theory}
\author{
{\large Satchidananda  Naik}
\\
  Harish-chandra Research Institute \\
 Chhatnag Road, Jhusi  \\
Allahabad-211 019, INDIA\\}
\begin{document}
\maketitle
\hspace*{\fill}
\hspace*{\fill}
\newcommand{\bee}{\begin{equation}}
\newcommand{\nn}{\nonumber}
\newcommand{\ee}{\end{equation}}
\newcommand{\ba}{\begin{array}}
\newcommand{\ea}{\end{array}}
\newcommand{\bea}{\begin{eqnarray}}
\newcommand{\eea}{\end{eqnarray}}
\newcommand{\ki}{\chi}
\newcommand{\eps}{\epsilon}
\newcommand{\pa}{\partial}
\newcommand{\lb}{\lbrack}
\newcommand{\Se}{S_{\rm eff}}
\newcommand{\rb}{\rbrack}
\newcommand{\de}{\delta}
\newcommand{\th}{\theta}
\newcommand{\rh}{\rho}
\newcommand{\ka}{\kappa}
\newcommand{\al}{\alpha}
\newcommand{\bt}{\beta}
\newcommand{\si}{\sigma}
\newcommand{\bsi}{\Sigma}
\newcommand{\vp}{\varphi}
\newcommand{\gm}{\gamma}
\newcommand{\gb}{\Gamma}
\newcommand{\om}{\omega}
\newcommand{\et}{\eta}
\newcommand{\gt}{ {g^2 T }\over{4 {\pi}^2}}
\newcommand{\qab}{{{\sum}_{a\neq b}}{{q_a q_b}\over{R_{ab}}}}
\newcommand{\omb}{\Omega}
\newcommand{\pr}{\prime}
\newcommand{\ra}{\rightarrow}
\newcommand{\nb}{\nabla}
\newcommand{\MSb}{{\overline {\rm MS}}}
\newcommand{\lnh}{\ln(h^2/\Lambda^2)}
\newcommand{\cz}{{\cal Z}}
\newcommand{\h}{{1\over2}}
\newcommand{\Lm}{\Lambda}
\newcommand{\inft}{\infty}
\newcommand{\bpa}{\bar {\partial}}  
\newcommand{\hth}{\hat {\theta}}
\newcommand{\hp}{\hat p}
\newcommand{\hb}{\hat b}
\newcommand{\hc}{\hat c}
\newcommand{\hbt}{\hat {\beta}}
\newcommand{\hgm}{\hat {\gamma}}
\newcommand{\hvp}{\hat {\varphi}}
\newcommand{\hlm}{\hat {\lambda}}    
\newcommand{\hv}{\hat v} 
\newcommand{\hq}{\hat q} 
\newcommand{\Lra}{\Longleftrightarrow}
\newcommand{\abschnitt}[1]{\par \noindent {\large {\bf {#1}}} \par}
\newcommand{\subabschnitt}[1]{\par \noindent
                                          {\normalsize {\it {#1}}} \par}
%-----------------------------------------------------------------------
% The definition below makes spaces e.g \skipp{3} makes 3 spaces
\newcommand{\skipp}[1]{\mbox{\hspace{#1 ex}}}
 
%
%
% various slashed symbols
%
%
%\newcommand\slash#1{\rlap{$#1$}/} % slashes a character
\newcommand\dsl{\,\raise.15ex\hbox{/}\mkern-13.5mu D}
    % this one can be subscripted
\newcommand\delsl{\raise.15ex\hbox{/}\kern-.57em\partial}
\newcommand\Ksl{\hbox{/\kern-.6000em\rm K}}
\newcommand\Asl{\hbox{/\kern-.6500em \rm A}}
\newcommand\Dsl{\hbox{/\kern-.6000em\rm D}} %roman D
\newcommand\Qsl{\hbox{/\kern-.6000em\rm Q}}
\newcommand\gradsl{\hbox{/\kern-.6500em$\nabla$}}
%--------
 %---------------------------------------------------------------
 %\vskip5.0cm
%\newpage
\begin{abstract} \normalsize
The anomalous Ward identity is derived for $N = 2$ SUSY Yang-Mills 
theories
, which is resulted out of Wrapping of $D_5$ branes on Supersymmetric two 
cycles.
From the Ward identity One obtains the Witten-Dijkgraaf-Verlinde-Verlinde
equation and hence can solve for the pre-potential. This way one avoids 
the
problem of enhancon which maligns the non-perturbative behaviour of the 
Yang-Mills
theory resulted out of Wrapped branes.
\end{abstract}
\vskip10.0cm
\newpage
\pagestyle{plain}
\setcounter{page}{1}
\abschnitt{1. Introduction}
Recently the gauge theory/gravity duality , which is commonly
known as Ads/CFT duality is extended to non-conformal pure
 ${\cal N}=1$ or ${\cal N}=2$  supersymmetric Yang -Mills (SYM)
theories \cite{Mald}. So far the perturbative behaviour of the ${\cal
N}=2$ SYM
is produced by this duality. The conventional folklore is that the
instantons which are responsible for the non-perturbative part of
the pre-potential are suppressed in the large N limit (since the
gauge gravity duality is valid only in the large N limit). Also the
non-perturbative strong coupling behaviour of SYM concerns the
dilaton which is plagued with
the  singularity of "enhancon" \cite{Polch}. Here we establish the 
anomalous super conformal Ward identity for ${\cal N}=2$ SYM in the
gravity dual picture. Further the super conformal Ward identity is
written as  Witten-Dijkgraaf-Verlinde-Verlinde (WDVV)\cite{WDVV} equation
from which one can obtain the exact pre-potential.
 \abschnitt{2. The Strategy}
We start with  type IIB little string theory e la' a collection of a
large number of $NS_5$ brane in the vanishing string coupling limit
which gives rise to $D =6$ SYM \cite{Seib}. Then we dimensionally reduce
two of its spatial world volume in such a way that we retain ${\cal N}=2$
SYM  in the low energy limit. The $NS_5$ brane has $SO(4)$ R-symmetry
as the normal bundle. When one identifies the $U(1)$ subgroup of the
$SO(4)$ R-symmetry with the $U(1)$ spin connection of the two cycle which
is compactified, one gets a covariant constant spinor and SUSY is
retained, which is commonly known as twisting \cite{Mald}.
This is called wrapping of $NS_5$ brane on a supersymmetric two cycles.
If the compact space is a two-sphere, then there will be no extra hyper-
multiplet and in the low energy limit i.e. in the scale much lower than
the radius of the sphere we will get pure  ${\cal N}=2$  SYM.
Thus it
amounts to consider a gauged $D = 7$ supergravity solution and then lift
it to get the solutions in ten dimensions. We use here the results of
\cite{gaunt,div} classical solutions of $D = 7$ gauged supergravity
which is amenable to ten dimensional string theory.
\bea
ds_{10}^2 &=& {\rm e}^{\Phi}\Bigg[ d x_{1,3}^2 +
\frac{z}{\lambda^2} \left( {d{\theta}}^2 +
\sin^2 {\theta} \,{d {\varphi}}^2 \right)
+\frac{1}{\lambda^2} \,{\rm e}^{2x} \,dz^2 \label{met10}\nn \\
&&~
+ \frac{1}{\lambda^2} \left(d {\theta}_1^2 +
\frac{{\rm e}^{-x}}{f(x)} \cos^2 {\theta}_1 \left(d
  \theta_{2} + \cos {\theta} \,d {\varphi}
\right)^2
+ \frac{{\rm e}^{x}}{f(x)} \sin^2 {\theta }_1\,d \theta_{3}^{2}
 \right)\Bigg]~~,
\eea
where the dilaton is
\bee
{\rm e}^{2\Phi} = {\rm e}^{2 z} \left[1 - \sin^2
{\theta}_1
~\frac{1+c \,{\rm e}^{-2 z}}{2  z} \right]~~
\ee
and
\bee
f(x) = {\rm e}^{x} \cos^2 {\theta }_1 + {\rm e}^{-x} \sin^2 {\theta}_1,
\ee
also
\bee
 {\rm e}^{-2x} = 1 - \frac{1+c \,{\rm e}^{-2 z}}{2  z}
\ee
where $\lambda$ is the  gauge  coupling constant of seven dimensional
gauged supergravity and $c$ is a parameter as the integration constant
of the classical solution. For $c\geq -1$ the range of the radial
variable is $z_0\leq z \leq \infty$ where $z_0$ is the solution
for $ {\rm e}^{-2x(z_0)} =0$. Here $\theta$ and $\varphi$ are the
angles of compact two-sphere with radius of compactification as
$\frac{z}{\lambda^2}$
and $\theta_1$, $\theta_2$ and $\theta_3$
are angles of transverse three-sphere.
The conservation of the RR-charge on
the transverse sphere $S_3$ fixes $\frac{1}{\lambda^2}= N\,g_s\,\alpha'$
for $N$ number of $D_5$ branes with string coupling $g_s$.
The $D_5$ brane action is given by
\bee
S = - \tau_5 \int d^6 \xi ~{\rm e}^{- \Phi}
\sqrt{- \det \left( G+ 2 \pi \alpha' F \right)} + \tau_5
\int\left(\sum_n C^{(n)}\wedge {\rm e}^{2 \pi \alpha' F}
\right)
\ee
where F is the world volume gauge field and $\tau_5$ is the brane
tension.
. The BPS condition is fixed from
the condition of the vanishing of the potential between two branes which
gives $\theta_1$ to be $\frac{\pi}{2}$. This condition makes the
transverse boundary of the D brane to be a two dimensional space
consisting $z$ and $\theta_3$ which will eventually the moduli space
of  ${\cal N}=2$  SYM.

We want to establish here the anomalous super conformal Ward
identity.  In the presence of gravity the trace anomaly
\bee
\langle {\theta}^\mu_\mu \rangle = \h
\frac{\bt(g)}{g^3}\left(F^a_{\mu\nu}\right)^2 ~ +
\frac{c(g^2)}{16{\pi}^2} \left(W_{\mu\nu\rho\si}\right)^2 ~ -
\frac{a(g^2)}{{16{\pi}^2}}\left({\tilde R}_{\mu\nu\rho\si}\right)^2
\ee
where $\bt(g)$ is the beta function of SYM, $a(g)$ and $c(g)$ are central
functions near the criticality, $W_{\mu\nu\rho\si}$ is the Weyl tensor
and ${\tilde R}_{\mu\nu\rho\si}$ is the dual of the curvature tensor.
However this relation can be extracted from the two point functions of
the energy momentum tensors\cite{ansl}. In Ref.\cite{gubkl}, it is shown
how to extract these functions from the absorption cross-section
of soft dilatons or gravitons  by the $D$ branes. The probability of
absorption is taken as
ratio of the flux  near $z_0$ to the in coming flux at very large z .
This gives
\bee
\si = \frac{N^4}{128{\pi}^3}\left(z -z_0\right)^2 {\omega}^3.
\ee
where $\omega$ is the frequency of the soft gluon. Here we see the
presence of $\left(z -z_0\right)^2$ which if we write
as in Ref.\cite{div}
 ${\rm e}^{z}=\rho$
we get a term $\left(\log\frac{\rho}{{\rho}_0}\right)^2$ in the
cross-section signaling the asymptotic freedom or logarithmic coupling.
Also we see the presence of enhancon when $z_0$ is zero. This gives
$\bt(g) = - \frac{N}{8\pi^2}g^3$. Similarly one can also calculate
from $U(1)$ R-current the chiral anomaly.
Combining trace anomaly, chiral anomaly and supertrace anomaly one can
write the Ward identity as \cite{howe}
\bee
2 {\cal F} - {\cal F(A) }'{\cal A} = \frac{N}{8\pi^2}\langle
tr{\psi}^2\rangle,
\ee
where ${\cal F}$ is the effective potential or the pre-potential,
${\cal A}$ is the chiral multiplet coupled to vector multiplet
in the  ${\cal N}=2$  SYM and $tr{\psi}^2$ is the anomaly multiplet
for example $trF^2$ will correspond to ${\theta}^{\mu}_{\mu}$.
Here the bosonic component of  ${\cal A}$ corresponds  to
$z-i{\theta}_3$ or in radial coordinate  $v {\rm e}^{{i\gm}}$. In
the broken phase
the branes will be distributed on a circle or ${\cal A}_i$ the eigen
values of ${\cal A}$ which are $U(N)$ matrices  will be distributed
on a circle. The
second part of eq.(8)
will read as ${\sum}_i\frac{\pa {\cal F}}{\pa {\cal A}_i}{\cal A}_i$.
 This
equation one can in principle solve in the large N limit and obtain the
exact pre-potential in this limit \cite{doug}.

 \end{document}